\documentclass[twocolumn,showpacs,superscriptaddress,aps,prb]{revtex4-1}

\usepackage{graphicx}
\usepackage{dcolumn}
\usepackage{bm}
\usepackage{epsfig}
\usepackage{epstopdf}

\begin{document}

\bibliographystyle{apsrev}

\title{High temperature thermoelectric response of double-doped SrTiO$_3$ epitaxial films}

\author{J. Ravichandran}
\email{jayakanth@berkeley.edu}
\affiliation{Applied Science and Technology Graduate Group, University of California, Berkeley, CA 94720}
\affiliation{Materials Sciences Division, Lawrence Berkeley National Laboratory, Berkeley, CA 94720}
\author{W. Siemons}
\affiliation{Department of Physics, University of California, Berkeley, CA 94720}
\author{D-W. Oh}
\affiliation{Department of Materials Science and Engineering, and  Materials Research Laboratory, University of Illinois, Urbana, IL 61801}
\author{J. T. Kardel}
\affiliation{Department of Materials Science and Engineering,University of California, Berkeley, CA 94720}
\author{A. Chari}
\affiliation{Department of Materials Science and Engineering,University of California, Berkeley, CA 94720}
\author{H. Heijmerikx}
\affiliation{Department of Physics, University of California, Berkeley, CA 94720}
\author{M. L. Scullin}
\affiliation{Department of Materials Science and Engineering,University of California, Berkeley, CA 94720}
\author{A. Majumdar}
\affiliation{ARPA-E, US Department of Energy, 1000 Independence Avenue, Washington, DC 20585}
\author{R. Ramesh}
\affiliation{Materials Sciences Division, Lawrence Berkeley National Laboratory, Berkeley, CA 94720} 
\affiliation{Department of Physics, University of California, Berkeley, CA 94720}
\affiliation{Department of Materials Science and Engineering,University of California, Berkeley, CA 94720}
\author{D. G. Cahill}
\affiliation{Department of Materials Science and Engineering, and Materials Research Laboratory, University of Illinois, Urbana, IL 61801}

\date{\today}

\begin{abstract}
SrTiO$_3$ is a promising $n$-type  oxide semiconductor for thermoelectric energy conversion. Epitaxial thin films of SrTiO$_3$ doped with both La and oxygen vacancies have been synthesized by pulsed laser deposition (PLD). The thermoelectric and galvanomagnetic properties of these films have been characterized at temperatures ranging from 300 K to 900 K and are typical of a doped semiconductor. Thermopower values of double-doped films are comparable to previous studies of La doped single crystals at similar carrier concentrations. The highest thermoelectric figure of merit ($ZT$) was measured to be 0.28 at 873 K at a carrier concentration of $2.5\times10^{21}$ cm$^{-3}$.
\end{abstract}


\maketitle

\section{\label{sec:level1}Introduction}

Thermoelectric phenomena deal with direct inter-conversion of heat flow and electrical power. Despite several attractive features of thermoelectric energy conversion, the low thermodynamic efficiency of thermoelectric materials limit their use to niche applications. The efficiency of these materials is directly related to the thermoelectric figure-of-merit, $Z$, given by $S^{2}\sigma/\kappa$, where $S$ is the thermopower or Seebeck coefficient, $\sigma$ and $\kappa$ are the electrical and thermal conductivity, respectively. Typically, obtaining a high figure-of-merit is a challenging task due to the coupling between the these parameters.\cite{disalvo99} Much research has been devoted to discovering new materials with higher efficiencies.\cite{snyder2008} Even though conventional semiconductors and intermetallics cater to a wide range of high figure-of-merit materials, the physics of thermoelectric transport in these materials is fairly well understood. On the other hand, transition metal oxides offer an alternative and interesting platform to study the physics of thermoelectricity, due to the $d$-band nature of the charge carriers and possibility of strong correlations. Two of the model systems for such oxides have been strontium titanate (SrTiO$_3$, STO)\cite{tokuda2001} and sodium cobaltate (Na$_x$CoO$_2$).\cite{terasaki1997}

STO is one of the best $n$-type thermoelectric oxides known to-date. It is a large band gap perovskite oxide and can be easily doped on both the A and B cationic sites or with oxygen vacancies to produce $n$-type carriers. The large orbital degeneracy of the Ti--\textit{d} conduction band in STO results in power factor comparable to state-of-the-art thermoelectric materials. The presence of the octahedral field of oxygen ions around the Ti splits the five fold degenerate \textit{d}--bands into t$_{2g}$ and e$_g$ bands. Unlike other dopants like La and Nb (no changes in lattice constants), incorporating oxygen vacancies in STO causes, both theoretically predicted\cite{wunderlich} and experimentally observed,\cite{gong} tetragonal distortion. Such a strain can alter both the position and degeneracies of the t$_{2g}$ and e$_g$ bands. In the film form, due to epitaxial strain, we can control the overall direction of this tetragonal distortion (out of plane). Hence, creating varying amounts of oxygen vacancies and La in epitaxial STO films can lead to filling controlled band engineering. The thermoelectric properties of STO have been studied in single crystals, \cite{tokuda2001, sohtajap2005} polycrystalline ceramics \cite{jliu2009, hmuta2003} and thin film forms\cite{sohtaapl2005} in the past. In the bulk single crystal or ceramic form, most investigations have employed only one kind of dopant on the A-site (e.g., La doping on Sr sites) or B-site (e.g., Nb doping on Ti sites) or oxygen vacancies; one notable exception is Liu \textit{et al.},\cite{jliu2009} who employed forming gas treatment on La doped STO ceramic samples to introduce oxygen vacancies. The introduction of oxygen vacancies in addition to cationic dopants is hard to accomplish in the bulk form. Conventionally, introduction of oxygen vacancies in the bulk form has been achieved by reduction treatments using vacuum or reducing gases. Such treatments do not generally create a uniform distribution of oxygen vacancies.  Using epitaxial thin film growth, it is possible to introduce oxygen vacancies with ease and in a controllable manner. In our particular case, we employ pulsed laser deposition (PLD) to  independently control the number of La and oxygen vacancies as dopants by changing the La doping in the target and the partial pressure of oxygen during growth. Even though we have not been able to determine the quantitative dependence of oxygen vacancy density on growth pressure, oxygen partial pressure during growth is qualitatively correlated with oxygen vacancy concentration. The terms oxygen partial pressure during growth and oxygen vacancy concentration shall be used interchangeably throughout the manuscript.
 
The measurement of thermoelectric transport properties on thin films present several challenges, particularly at high temperatures. Often, there is a very limited choice of substrates for the growth of epitaxial films due to stringent requirements such as low lattice mismatch, stability under high temperatures and high vacuum.  In this case, the substrates need to remain electrically insulating, during the growth and measurements, to ensure easy and direct measurement of film's electrical properties without any contribution from the substrate.\cite{scullin2008} Furthermore, there are only a handful of techniques\cite{cahill1987, cahill2004} to measure thermal conductivity of epitaxial thin films reliably without removing the film from the substrate. Such issues make thermoelectric measurements on thin films
difficult, particulary at high temperatures.

In this article, we report measurements of all the three thermoelectric parameters on epitaxial thin films of SrTiO$_3$, doped with La and oxygen vacancies, in the temperature range $300 < T< 900$ K. We also report the galvanomagnetic properties such as carrier concentration and Hall mobility and compare the data with existing literature. 

\section{\label{sec:level1}Experimental Methods}

 Thin films of Sr$_{1-x}$La$_x$TiO$_{3-\delta}$ (150 nm thick) were grown by PLD from dense, polycrystalline ceramic STO targets each containing 0, 2, 5, 10 or 15 atomic \% of La substituted for Sr. We chose to grow the films on LSAT (001) substrates ((LaAlO$_3$)$_{0.3}$-(Sr$_2$AlTaO$_6$)$_{0.7}$), due to the close lattice matching with STO (a$_{STO}$ = 3.905 \AA ~and pseudocubic a$_{LSAT}$ = 3.87 \AA) and electrically insulating nature under high temperatures and high vacuum. The substrate  had dimensions of 5 mm x 5 mm x 0.5 mm. The growth of the films was carried out at 850$^\circ$C in an oxygen partial pressures ranging from 10$^{-1}$--10$^{-8}$ Torr using a 248 nm KrF excimer laser with a fluence of 1.5 J/cm$^2$ at a repetition rate of 2--8 Hz. The temperature of 850$^\circ$C is measured on the heater; the substrate temperature during growth is lower. X-ray diffraction (XRD) was carried out on these films with a Panalytical X'Pert Pro thin film diffractometer using Cu K$\alpha$ radiation. Thicknesses of the films were determined using X-ray reflectivity, Rutherford backscattering spectrometry (RBS) and/or {\it in-situ} reflection high energy electron diffraction (RHEED). The stoichiometry and chemical composition of the films were characterized using RBS. 

 All transport measurements were carried out in the van der Pauw geometry. Triangular metal contacts of side 1 mm (15 nm Ti/100 nm Pd) were deposited on the corners of the films using electron beam evaporation. Galvanomagnetic measurements were carried out in air using a home-built apparatus. The apparatus is equipped to measure properties over 300 - 900 K using a 1 Tesla electromagnet. Thermoelectric measurements were performed in an evacuated appartus maintained at 10$^{-4}$--10$^{-5}$ Torr. Thermopower measurements were performed using K-type thermocouples. Before the measurements, the thermocouples were thermally anchored to the corners of the sample on the metal contacts using silver paint. Temperature differences across the sample, with a step of 2 K and a maximum value of 6 K (both positive and negative), were used to extract the thermopower values. The alumel leads were used to measure the thermoelectric voltage generated across the sample due to the applied temperature gradient. The slope of the thermoelectric voltage against the temperature difference was corrected with the known thermopower values of alumel\cite{bentley} at a given average temperature to obtain the thermopower of our samples. Thermal conductivity measurements were performed by time domain thermoreflectance technique (TDTR).\cite{cahill2004} All pressures referred in the manuscript are oxygen partial pressures unless specified otherwise. In all the cases above, the transport measurements showed little or no hysteresis on cycling the samples several times, either in air or vacuum. This suggests that the oxygen vacancies are kinetically stable in the temperature range and time scales at which the measurements were carried out. 

\section{\label{sec:level1}Results and Discussion}

\subsection{\label{sec:level2}Structural and chemical characterizations}

\setlength{\epsfxsize}{1\columnwidth}
\begin{figure}[t]
\epsfbox{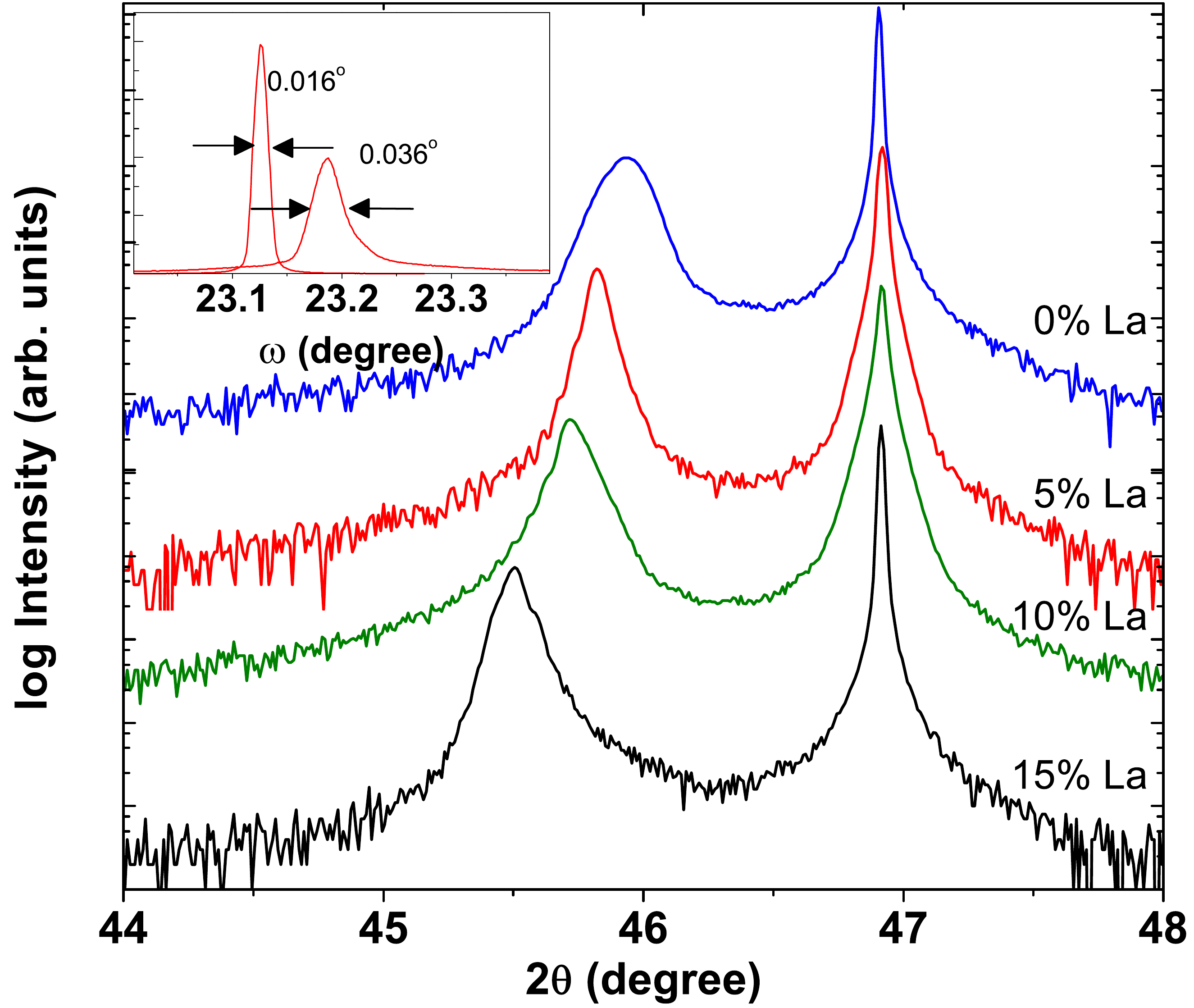}
\caption{\label{Fig:fig1} The out-of-plane XRD patterns for films grown with different La doping at 10$^{-8}$ Torr. Each curve is labeled by the La doping level. The inset shows a comparison of the (002) rocking curves of the substrate and a film with 5\% La doping grown at 10$^{-8}$ Torr. The rocking curve FWHM for the substrate and the film are 0.016$^\circ$ and 0.036$^\circ$.
}
\end{figure}

 The structural properties of the films were characterized using x-ray diffraction (XRD). All the films were single phase, without any trace of secondary phases. Representative XRD patterns for the (00$l$) planes of films grown at 10$^{-8}$ Torr with various La doping are shown in Fig.~\ref{Fig:fig1}. The patterns show that there is a clear increase in the $c$-axis lattice parameter with increasing La concentration. A similar effect was observed for increasing oxygen vacancy concentration (decreasing growth pressure) for a fixed La concentration. There are several possible reasons for the increase in \textit{c}-axis lattice constant in STO, such as non-stoichiometry\cite{brooks}, theoretically predicted tetragonal distortion caused by the presence of oxygen vacancies/vacancy clusters,\cite{wunderlich, wluo} filling of the carriers in the conduction band,\cite{walle} and to a lesser extent, in-plane bi-axial strain caused by the substrate (films strain relax partially or completely at thicknesses $\sim$ 150 nm). In the bulk crystal form, the lattice constants increase isotropically by less than 0.1\% for upto 20\% La doping and retains the cubic structure\cite{sunstrom}. This suggests that the lattice expansion observed in our case should be a combination of one or more of the above mentioned reasons (we observed partial strain relaxation in the in-plane lattice parameters for all films). The inset shows the comparison of the rocking curve for the substrate and the film with 5\% La doping grown at 10$^{-8}$ Torr. The FWHM for the rocking curve of film peak was 0.036$^\circ$ as compared to the substrate with 0.016$^\circ$. RBS measurements confirmed the chemical composition and homogeneity of the films within the limits of experimental errors. RBS channeling experiments (data not shown) established good crystallinity of the films, corroborating the XRD rocking curve data.

\subsection{\label{sec:level2}Thermoelectric properties}

\setlength{\epsfxsize}{1\columnwidth}
\begin{figure}[t]
\epsfbox{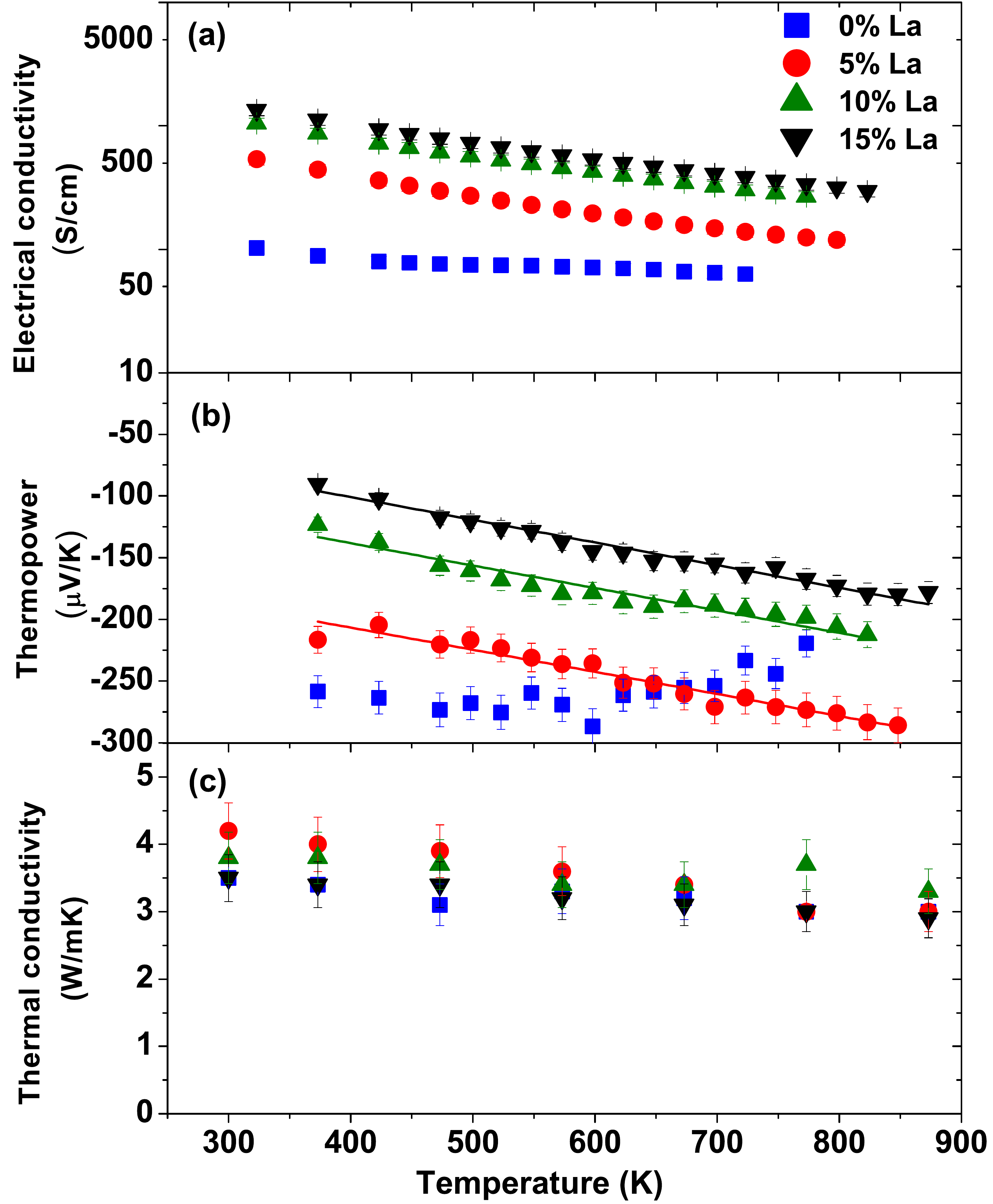}
\caption{\label{Fig:fig2} The thermoelectric properties, namely (a) electrical conductivity, 
(b) thermopower and (c) thermal conductivity, are shown for films grown with different La 
doping (0, 5, 10 and 15 \%) at 10$^{-8}$ Torr. Electrical conductivity for all samples decreased with increasing temperature and the thermopower data was fit by a linear temperature depedance for a finite La doping.}
\end{figure}

 We discuss the thermoelectric properties of the samples grown at 10$^{-8}$ Torr, whose structural properties were discussed earlier. The thermoelectric properties of the films grown at 10$^{-8}$ Torr with various La doping are shown in Fig.~\ref{Fig:fig2}. The conductivity of the films decreased with increasing temperature, characteristic of a heavily doped semiconductor. The electrical conductivity also increased with an increasing La doping concentration. The thermopower values for the 0\% La doping sample shows a non-linear temperature dependence with a maximum at 500 -- 600 K. Interestingly, the thermal band gap calculated using the relation E$_g$ = 2eS$_{max}$T$_{max}$ \cite{goldsmid99} for this case is about 0.3 eV. Several theoretical predictions\cite{wunderlich, wluo} and experimental observations\cite{mochizuki,jravi} have established that the introduction of oxygen vacancies can create a defect band ~0.3 eV below the conduction band of STO. It is possible that we are observing the gap between the oxygen vacancy defect band and the conduction band for this sample. For finite La doping, thermopower shows a linear increase with temperature, similar to the observation of Muta \textit{et al.}\cite{hmuta2005} Even though we observe a linear temperature dependence of thermopower, it is hard to extract useful information on Fermi energy, as the doping regime is in a transition from non-degenerate to degenerate for these carrier concentrations. This transition will be later discussed based on the galvanomagnetic properties. 

\setlength{\epsfxsize}{1\columnwidth}
\begin{figure}[t]
\epsfbox{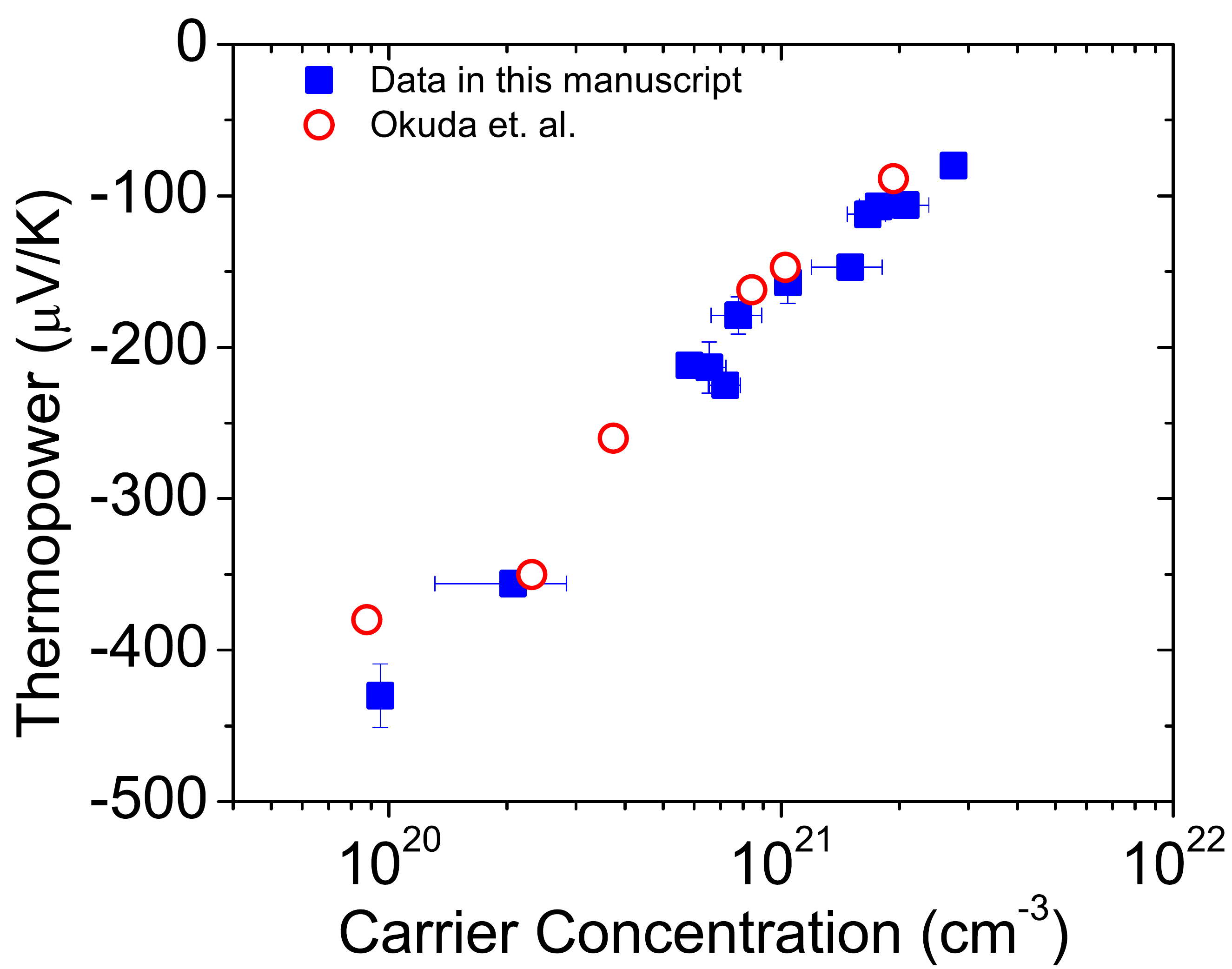}
\caption{\label{Fig:fig5} Thermopower values at 300 K for various La doping and growth pressures
are plotted against their respective carrier concentrations. The open circles are data from Okuda
\textit{et al.}\cite{tokuda2001} showing good agreement with our data.}
\end{figure}

 The thermal conductivity values for all the samples decreases from $\sim$ 4--5 W/mK at 300 K to $\sim$ 3 W/mK at 873 K. 
The measured room temperature thermal conductivity is a factor of 2--3 lower than the reported single crystalline thermal conducitivity,\cite{sohtaapl2005} possibly due to large number of defects in the film and so, close to room temperature the thermal transport is dominated by defects. The thermal conductivity values at 800--900 K are similar to the single crystalline values, suggesting that the dominant scattering process is possibly not due to defects but due to Umklapp scattering as observed in single crystals. All thermal conductivity measurements reported here are for the cross-plane direction of the thin films. Nevertheless, due to the near cubic structure of these heavily doped STO films (bulk STO is cubic), it is reasonable to assume that the thermal conductivity anisotropy is negligible.

 To put our thermopower data in perspective, we have plotted our thermopower values at 300 K over the whole carrier concentration range with the values of Okuda \textit{et al.}\cite{tokuda2001} in Fig.~\ref{Fig:fig5}. It is important to note that there is little or no effect of oxygen vacancies in increasing the thermopower, contrary to what we expected due to band engineering. If the tetragonal distortion is not strong enough to alter the crystal field split bands, we may not see the effect of band engineering at all, particularly with increasing temperatures, due to thermal broadening the effect of tetragonal distortion shall decrease. Any estimate of the energy scales and concentration of vacancies required to see a desirable effect will need in-depth first principle calculations and is beyond the scope of this current study.

\begin{figure}[t]
\epsfbox{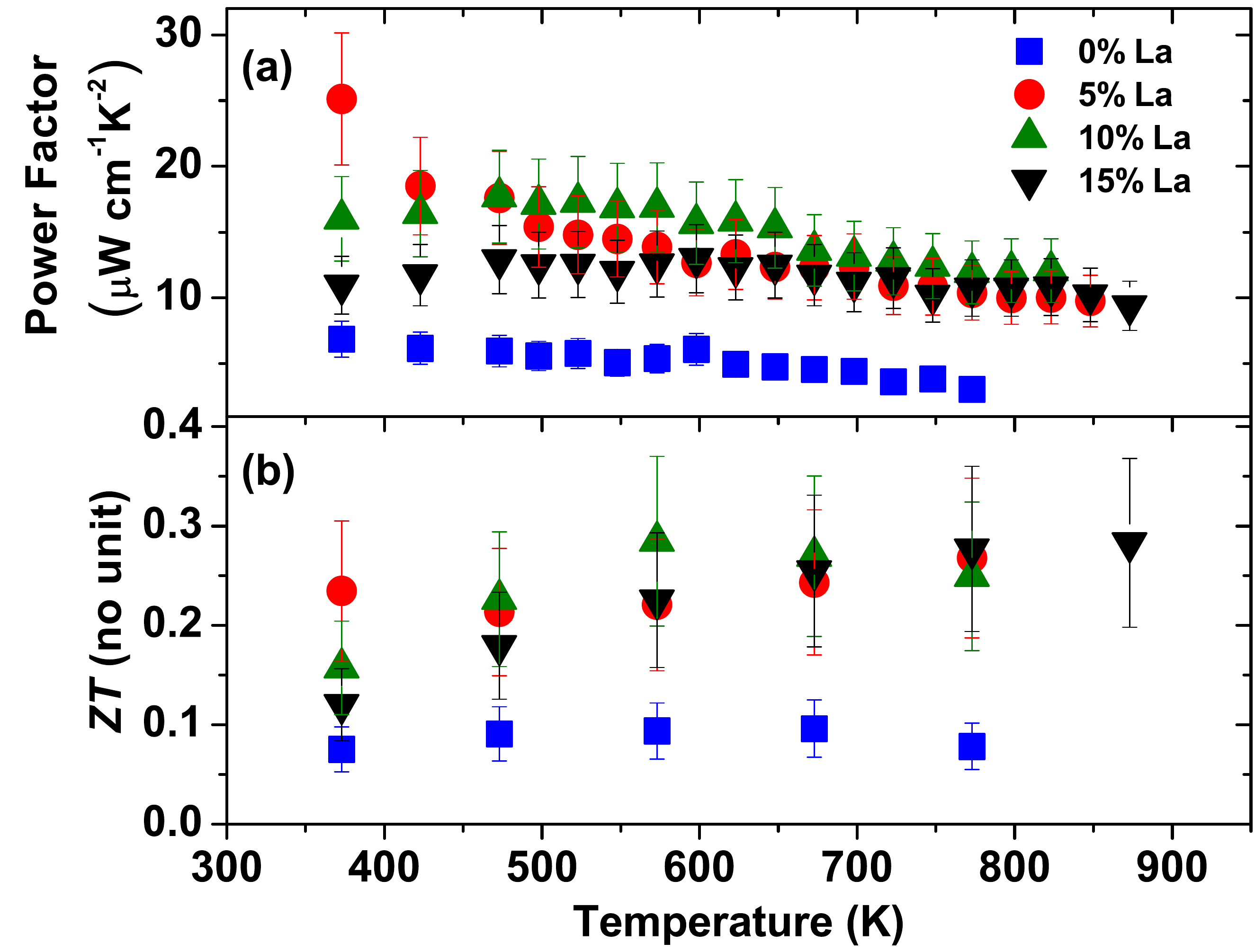}
\caption{\label{Fig:fig3} The power factor (\textit{S$^{2}\sigma$}) and \textit{ZT} for the films shown in 
Fig.~\ref{Fig:fig2}.}
\end{figure}

 The derived power factor ($S^{2}\sigma$) and the dimensionless figure of merit, $ZT$ are shown in Fig.~\ref{Fig:fig3}. At room temperature 5\% La doped sample showed the highest $ZT\sim0.18$ but the highest $ZT$ of 0.28 was observed at 873 K for the 15\% La doped sample at 873 K. Recently, several authors have reported high $ZT$ in bulk STO using forming gas reduction with La doping (ZT=0.21 at 750 K),\cite{jliu2009} Dy and La co-doping (ZT=0.36 at 1045 K),\cite{wang2010} 
and Dy and Nb co-doping (ZT=0.24 at 1273 K).\cite{tinh2009} Our results are as high as the values reported by these groups in the reported temperature range. 

\setlength{\epsfxsize}{1\columnwidth}

\subsection{\label{sec:level2}Galvanomagnetic properties}

\setlength{\epsfxsize}{1\columnwidth}
\begin{figure}[t]
\epsfbox{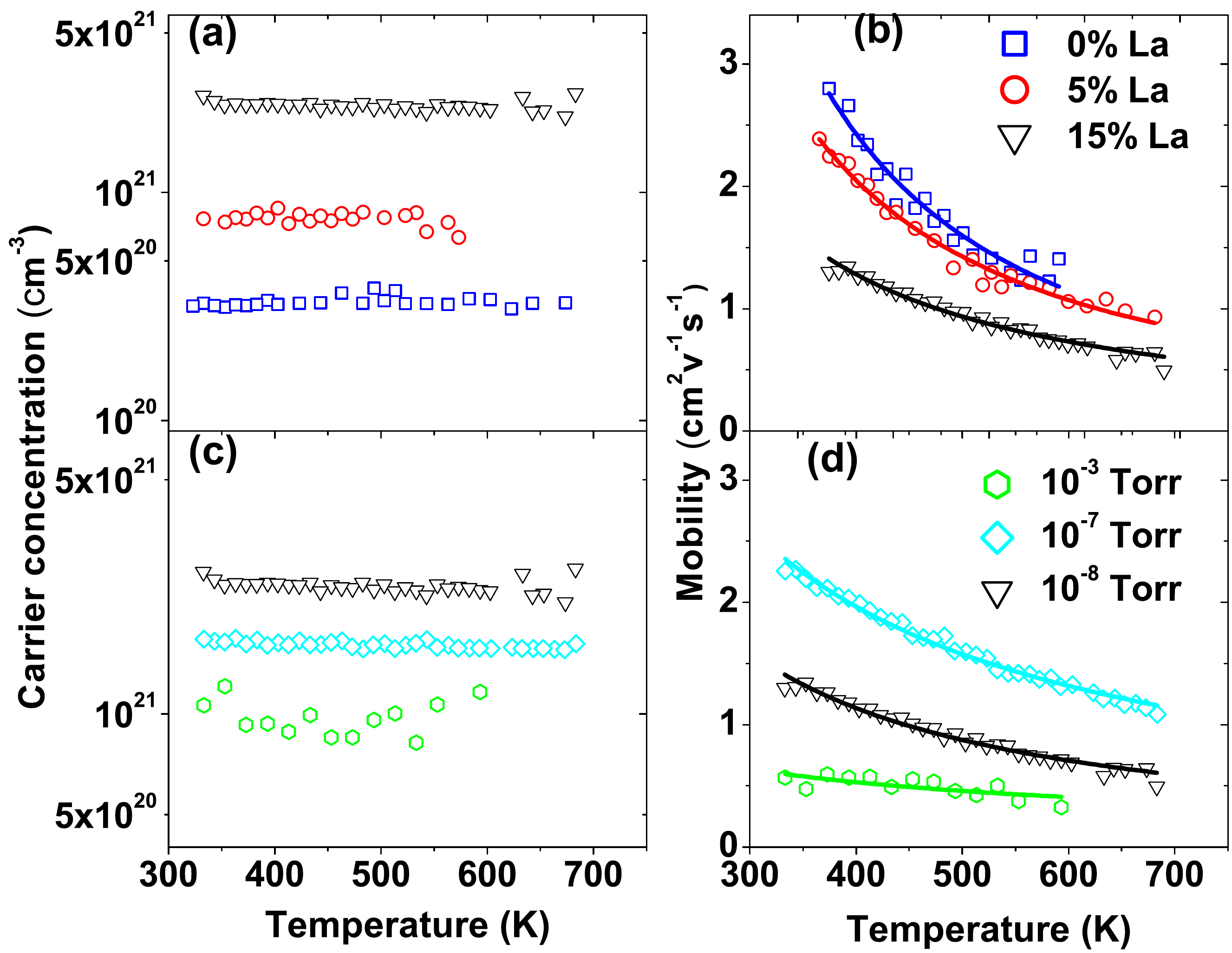}
\caption{\label{Fig:fig4} The carrier concentration and Hall mobility for (a) films grown at 10$^{-8}$ Torr with different La doping such as 0, 5, 15 \% and (b) films grown with 15\% La doping grown at different pressures such as 10$^{-3}$, 10$^{-7}$ and 10$^{-8}$ Torr. The mobility data was fit using power law as described in the text.}
\end{figure}

 Fig.~\ref{Fig:fig4} shows the derived carrier concentration and Hall mobility for two sets of samples. The samples had either the same growth pressure (10$^{-8}$ Torr) with different La doping (0, 5 and 15\%) as shown in Fig.~\ref{Fig:fig4}(a)\&(b)) or constant La doping (15\%) with different growth pressures of 10$^{-3}$, 10$^{-7}$ and 10$^{-8}$ Torr as shown in Fig.~\ref{Fig:fig4}(c)\&(d)). In all cases, the carrier concentration remained constant over the measured temperature range. No hysteresis was observed in all the measured properties over 2-3 measurement cycles, suggesting excellent stability of oxygen vacancies present in the films. It is evident that oxygen vacancies play a significant role in modulating the carrier density in the films, as much as the La doping does. 

In all cases the temperature dependent mobility was fitted using a power law model as described below.
\begin{equation}
\mu = AT^{-n}
\label{Eq:mobility}
\end{equation}
where, A is the prefactor and n is the power law exponent. The power law exponent reported in the literature varies from 1.5--2.7 over a temperature range of 100--1300 K. Tufte \textit{et al.}\cite{tufte1967} reported an exponent of 2.7 around 100--400 K. Uematsu \textit{et al.}\cite{uematsu} observed an exponent of 2.0 over a temperature range of 500--1000 K. Moos \textit{et al.}\cite{moos1996} noted a change in the exponent from 2.7 close to room temperature to 1.6 above 1000 K. Unlike these observations, Ohta \textit{et al.}\cite{sohtajap2005} reported an exponent of 1.5 over 300--1000 K and correlated the behavior to phonon scattering as observed in classical semiconductors\cite{vining}. 

\setlength{\epsfxsize}{1\columnwidth}
\begin{figure}[t]
\epsfbox{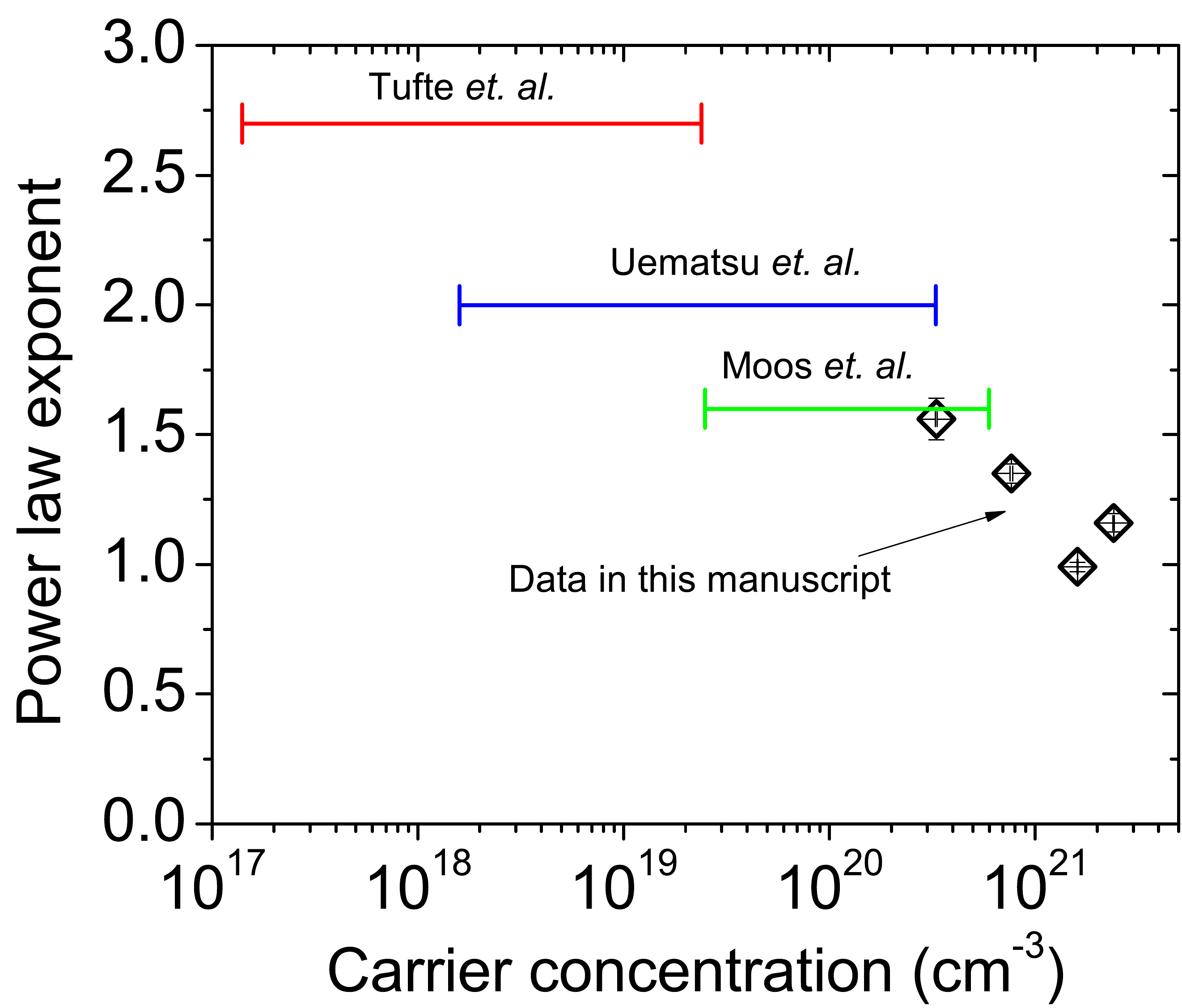}
\caption{\label{Fig:fig6} The mobility power law exponent as a function of carrier concentration. The represented carrier concentration value is averaged over the whole temperature range for our data. The exponents reported in the literature are also given for reference.}
\end{figure}

Previous studies found no dependence of the exponent on the carrier concentration. We observed a systematic change in the power law exponent as a function of carrier concentration in our samples as depicted in Fig.~\ref{Fig:fig6}. We observe a change in the power law exponent from $\sim$ 1.5 to $\sim$ 1.0 as the carrier concentration increases. It is important to note that the carrier concentrations of the samples used in the prior work were atleast an order of magnitude lower than our highest carrier concentration. In a classical semiconductor, the power law exponent for acoustic/optical phonon scattering changes from 1.5 to 1.0 with a change from non-degenerate doping to fully degenerate doping.\cite{vining}

\section{\label{sec:level1}Summary}

 In summary, we have studied the thermoelectric and galvanomagnetic response of thin films of La and oxygen vacancy doped SrTiO$_3$. Even though the films show very similar electrical properties compared to bulk La or oxygen vacancy doped STO crystals, the thermal properties differ in the form of low thermal conductivity close to the room temperature. At the heavy doping limit, as was investigated in this manuscript, we found no effect of the defect band or band engineering due to the oxygen vacancies on thermopower but oxygen vacancies do provide additional itinerant carriers to the conduction band. Most notably, we have performed all thermoelectric measurements on thin films of STO and the highest figure merit achieved was $\sim$ 0.28 $\pm$ 0.08 at 873 K.

\begin{acknowledgments}
The research at Berkeley was supported by the Division of Materials Sciences and Engineering, Office of Basic Energy Sciences, U.S Department of Energy under the contract no. DE-AC02-05CH11231. The research at Illinois was supported by a DOE grant DEFG02-07ER46459. JR acknowledges the support from Link foundation. WS acknowledges the Netherlands organization of scientific research (NWO). The authors gratefully acknowledge the assistance of Kin Man Yu in the RBS measurements, Yee Kan Koh in thermal conductivity measurements, Martin Gajek in Hall measurements and discussions with Subroto Mukerjee and Jay Sau. 
\end{acknowledgments}

\end{document}